\documentclass[a4paper,11pt]{article}
\pdfoutput=1 

\usepackage{jheppub} 

\usepackage[T1]{fontenc} 
\usepackage{amsmath}
\usepackage{amsfonts}
\usepackage{amssymb}
\usepackage{hepunits}
\usepackage[font=small,skip=0pt]{caption}
\usepackage{float}
\usepackage[utf8]{inputenc}
\usepackage[colorinlistoftodos]{todonotes}

\title{\boldmath Decays of Long-Lived Relics and Their Signatures at IceCube }

\author[1]{Kim~ V.~ Berghaus,\note{corresponding author}}
\author{Melissa~ D.~ Diamond,}
\author{D.~ E.~ Kaplan}

\affiliation{Department of Physics $\&$ Astronomy,\\The Johns Hopkins University, Baltimore, MD 21218, USA}

\emailAdd{kbergha1@jhu.edu}
\emailAdd{mdiamon8@jhu.edu}
\emailAdd{david.kaplan@jhu.edu}

\abstract{We consider long-lived relic particles as the source of the PeV-scale neutrinos detected at the IceCube observatory over the last six years. We derive the present day neutrino flux, including primary neutrinos from direct decays, secondary neutrinos from electroweak showering, and tertiary neutrinos from re-scatters off the relic neutrino background. We compare the high-energy neutrino flux prediction to the most recently available datasets and find qualitative differences to expected spectra from other astrophysical processes. We utilize electroweak corrections to constrain heavy decaying relic abundances, using measurements impacted by electromagnetic energy injection, such as light element abundances during Big Bang nucleosynthesis, cosmic microwave background anisotropies, and diffuse $\gamma$-ray spectra. We compare these abundances to those necessary to source the IceCube neutrinos and find two viable regions in parameter space, ultimately testable by future neutrino, $\gamma$-ray, and cosmic microwave background observatories. }

\begin{document} 
\maketitle
\flushbottom

\section{Introduction}

The IceCube detector, located in the Antarctic ice layer, is sensitive to neutrino energies ranging from $10$ - $10^{10}\,$GeV \cite{Halzen:2010yj}. Over its six year run, IceCube has detected several neutrinos in the energy range $30\,$TeV$\,$-$\,10\,$PeV \cite{Aartsen:2013jdh,Kopper:2015vzf,Kopper:2017zzm,Aartsen:2016xlq,Aartsen:2014gkd}. The measured neutrino flux in this range is significantly larger than that expected from the atmospheric neutrino background \cite{Aartsen:2013jdh,Kopper:2015vzf,Kopper:2017zzm,Aartsen:2016xlq}. This suggests an alternative source with a significance of at least $7\sigma$ \cite{Proceedings:2016rni}. Previously, no statistically significant correlation between the direction of origin of the detected neutrinos and any known high energy $\gamma$-ray sources existed, suggesting an isotropic extra-galactic source \cite{Aartsen:2016oji}. Recently however, multi-messenger astrophysics linked one $290\,$TeV neutrino to a flaring blazar \cite{IceCube:2018dnn}. More data is necessary to determine whether blazars can explain the highest energy events. Other possible astrophysical sources such as Supernova remnants (SNRs), star forming regions, Fermi bubbles, and active galactic nuclei (AGNs), have also been considered in the past \cite{Essey:2010er,Kalashev:2013vba,Stecker:2013jfa,Tamborra:2014xia,Murase:2013rfa,Lunardini:2013gva,Dermer:2014vaa,Padovani:2014bha,Murase:2014foa}. Beyond the standard model physics (BSM) explanations have been investigated with regard to heavy decaying dark matter (see, for example, \cite{Bai:2013nga,Esmaili:2013gha,Rott:2014kfa,Esmaili:2014rma,Murase:2015gea,Ko:2015nma,Cohen:2016uyg,Dev:2016qbd,Bhattacharya:2017jaw,Borah:2017xgm,Sui:2018bbh,Aartsen:2018mxl}). However, many models of decaying dark matter as a source of the IceCube neutrinos are highly constrained because they are predicted to produce $\gamma$-rays in excess of current measurements \cite{Esmaili:2015xpa,Cohen:2016uyg,Aartsen:2018mxl}. In this paper, we explore the experimental signature of a heavy relic directly decaying to neutrinos, sourcing an isotropic extra-galactic high-energy neutrino flux. We focus on lifetimes that are shorter than the age of the universe. We examine whether this high-energy neutrino flux can fit the excess events seen between $250\,$TeV$\,$-$\,10\,$PeV. We show that many constraints imposed by $\gamma$-ray observations can be avoided under this set of assumptions. 

Recently, electroweak corrections at energy scales well above the electroweak (EW) scale have drawn considerable attention \cite{Ciafaloni:2010ti,Chen:2016wkt,Sarkar:2001se,Barbot:2002ep}. For high-energy scattering and decays, the EW effects significantly impact phenomenology by producing higher multiplicity final states. Different implementation strategies have been explored with regards to heavy decaying DM \cite{Ciafaloni:2010ti, Ciafaloni:2011sa,Cavasonza:2014xra}. In our analysis we implement a fixed order EW shower.  We use the results of the shower to predict a neutrino spectrum and fit it to that detected at IceCube. We also explore how the decaying relic model is constrained by its impact on  light element abundances, CMB anisotropies, and diffuse $\gamma$-ray spectra, after including the EW shower effects. 

A long lived relic has been considered previously as a source for the IceCube neutrinos, and analyzed up to redshifts of $z = 1000$ \cite{Ema:2013nda,Ema:2014ufa}. We extend this range by including neutrinos arising from re-scatterings off the relic neutrino background in our analysis. Our inclusion of EW corrections further changes the qualitative features of the neutrino flux today, leading us to conclude that EW corrections are a necessary part of an accurate forecast.

\section{Models}

In this paper, we consider two models in which our relic, $X$, directly decays to neutrinos.  In our analysis, the PeV-scale neutrinos observed at IceCube are assumed to come from these direct decays. Naively, one may wish to consider a toy-model decay: $X \to \overline{\nu}\nu$ \cite{Esmaili:2012us}. However, implementing an EW shower highlights the inconsistency of this treatment. At ultra-high energies, the final state radiation includes many soft $W$'s,  which turn charged leptons into neutral ones and vice versa. This leads to the production of roughly the same amount of neutral and charged leptons for center-of-mass (COM) energies far beyond the EW scale. This is a side effect of unbroken isospin in the high-energy limit. Model-independently, this implies that any high-energy neutrino spectrum sourced directly from a heavy relic decay will be accompanied by a spectrum of electromagnetically interacting particles, which will carry roughly the same amount of energy as the neutrino spectrum. 

At energy scales much above the EW scale, Sudakov logarithms contribute to higher-multiplicity final states. These corrections grow logarithmically as the mass increases. Effectively this leads to the production of EW jets. To quantify the neutrino spectrum arising from these jets, we implement a fixed order EW shower. The qualitative features of the EW jets are model independent, as any heavy particle that decays to neutrinos will also radiate gauge bosons. To zeroth order, this effect takes a delta function centered around $\frac{M_X}{2}$, and smears it towards lower energies. The energy lost by the neutrinos is carried away by gauge bosons, which themselves can decay into neutrinos, and contribute to the neutrino spectrum at lower energies.  We describe the implementation of the EW shower in detail in Appendix \ref{EW shower}.

We consider two benchmark models that produce neutrinos through direct decays while remaining consistent with the isospin structure dictated by the Standard Model. We do not study a specific production mechanism for the heavy relic abundance, and assume it is cold. We note that inflationary dynamics can trivially produce such a particle during the reheating period \cite{Kane:2015jia}. Model-dependent constraints on these production mechanisms exist based on measurements such as isocurvature; however, these are not stringent enough to rule out the small abundance of decaying relics necessary to source the IceCube neutrinos \cite{Kainulainen:2016vzv, Heikinheimo:2016yds}.

\subsection{Model I: Heavy Scalar $X_1$}
We consider a heavy scalar $X_1$, that couples to the standard model lepton doublets $L^i$. Here $i = 1,2,3$ indexes the generation. For simplicity, we assume flavor universality:
\begin{equation}\label{lag2} 
\mathcal{L}_1 = \frac{1}{2} \partial_{\mu} X_1 \partial^{\mu} X_1  - \frac{1}{2}M_X^2 X_1^2 + g_1 {L^i}^{\dagger} \overline{\sigma}^{\mu} \partial_{\mu} L^i X_1
\end{equation}
The zeroth order decays are given by: 
\begin{equation} \label{decay1}
\begin{split}
& X_1 \to \ell^+ \ell^-   \\
& X_1 \to \overline{\nu_\ell}\nu_\ell
\end{split}
\end{equation} 
The ratio of branching ratios is essentially $1:1$ at tree level. We will refer to the above decay model I as $X \rightarrow \overline{\nu} \nu$.
\subsection{Model II: Heavy Fermion $X_2$} 
In our second model we consider a heavy Dirac fermion, that couples to the standard model lepton doublets ($L^i$) and Higgs doublet ($\phi$). 
\begin{equation}\label{lag1} 
\mathcal{L}_2 = \frac{i}{2} X_2^{\dagger} \overline{\sigma}^{\mu}\partial_{\mu} X_2 - \frac{1}{2} M_X \left(X_2 X_2 + X_2^{\dagger} X_2^{\dagger} \right) + g_2 \phi^{\dagger} L^i X_2 + g_2^{\dagger} {L^i}^{\dagger} \phi X_2^{\dagger}
\end{equation}
We assume relic and its anti-particle have the same number density. The zeroth-order decays of $X_2$ are given by: 
\begin{equation} \label{decay21}
\begin{split}
& X_2 \to \ell W,      \\
& X_2 \to \nu_\ell Z/h,     \\
\end{split}
\end{equation} 
Again, the decays to $W^\pm, Z, h$ have  equal branching ratios at tree-level in the high mass limit. We will refer to the above decay model II as $X \rightarrow V \ell$. 

\section{The Neutrino Spectrum}

\subsection{Derivation of the Present-Day Neutrino Flux}
To derive the shape of the differential flux today we extend the analysis performed in \cite{Ema:2013nda}. We consider a number density of cold heavy relic $X$s that decay with a given lifetime $\tau_X$:
\begin{equation}
n_X(t) = n_{X,0}(t) \mathrm{e}^{-\frac{t}{\tau_X}}, 
\end{equation}
where $n_{X,0}(t)$ is the number density in the limit $\tau_X\rightarrow\infty$.  For any given decay, high-energy neutrinos are injected into the thermal bath. The maximum possible energy is set by the mass of the heavy relic: $E_{{\text{max}}}= \frac{M_X}{2}$. The fractional energy distribution $f_{E_{\text{max}}}(x)$ of these neutrinos is determined by the EW shower, where $x = \frac{E}{E_{\text{max}}}$ and $E$ is the injection energy of the neutrino. This decay gives rise to the following source term:
\begin{equation}
S_{\text{dec}}(t,E) = n_X(t) \frac{1}{4 \pi \tau_X} \frac{f_{E_{\text{max}}}\left(\frac{E}{E_{\text{max}}}\right)}{E_{\text{max}}}  
\end{equation} 
Depending on when they were produced, the neutrinos may free-stream or scatter off the relic neutrino background. The cross sections for all relevant (anti-)neutrino-(anti-)neutrino scattering processes are listed in \cite{Ema:2013nda}. The total scattering rate is determined by the thermally averaged cross section: 
\begin{equation}
\Gamma_{\text{tot}} = n_{\text{BG}} \langle \sigma_{\text{tot}} v_{\text{rel}} \rangle
\end{equation}
In the massless neutrino limit, the relative velocity simplifies to: $v_{\text{rel}} = \frac{s}{2 E k}$, where $s$ is the squared COM energy and $k$ is the energy of the relic background neutrino. The scattering rate can then be written as \cite{Ema:2013nda}:
\begin{eqnarray}
\Gamma_{\text{tot}}(t,E) &=& \frac{1}{16 \pi^2 E^2} \int dk \frac{1}{1+e^\frac{k}{T_{\nu}(t)}} \int_0^{4 k E} ds s \sigma_{\text{tot}} (s) \\
&=& \frac{T_{\nu}(t)}{\pi^2} \int dk k \ln \left(1 + e^{-\frac{k}{T_{\nu}(t)}}\right) \sigma_{\text{tot}}(s = 4 k E) 
\end{eqnarray}
where the second line is achieved via integration by parts.  Neutrinos that scatter off the relic neutrino background, at the COM energies we consider, may produce two energetic neutrinos, two charged leptons, or two quarks. We define $\Gamma_{\nu}$ and $\sigma_{\nu}$ as the scattering rate and cross-section for $2\rightarrow 2$ neutrino scattering. We account for this re-injection of neutrinos by adding an additional source term. This is sometimes referred to as a tertiary source term \cite{Boudaud:2014qra}. 
\begin{equation} \label{source2}
S_{\text{ter}}(t,E) =  \int_E^{\infty} \frac{1}{\sigma_{\nu}(t,E')}\frac{d\sigma_{\nu}}{dE} \Gamma_{\nu}(t,E') \Phi(t, E') dE' 
\end{equation}
where $\Phi(t,E')$ is the differential neutrino flux defined in terms of the neutrino number density $n_\nu(t) = \int dE' \Phi(t,E')$, and $E$ is the scattered neutrino energy.

We simplify equation \eqref{source2} by rewriting the differential cross section in terms of the injection energy, $E'$, and the fractional scattered energy $ y = E/E'$. 
\begin{equation} \label{prox}
\frac{1}{\sigma_{\nu}} \frac{d\sigma_{\nu}}{dE} \approx \frac{1}{E'} \frac{1}{\sigma_{\nu}} \frac{d\sigma_{\nu}}{dy} \equiv \frac{1}{E'} g(y)
\end{equation}
We can make the approximation in equation \eqref{prox} because, for large boosts ($\gamma > 100$), $g(y)$ becomes independent of $E'$. We derive $g(y)$ by boosting the relevant differential cross sections from the COM-frame to the laboratory frame:
\begin{equation}
\frac{d\sigma_{\nu\nu}}{d\Omega_{\text{COM}}} \propto 1
\end{equation}
\begin{equation}
\frac{d\sigma_{\bar{\nu}\nu}}{d\Omega_{\text{COM}}} \propto (1+\cos \theta)^2
\end{equation}
Defining separate functions $g(y)$ for each scattering independently -- for neutrino-neutrino scattering (and its conjugate scattering), $g_{\nu\nu}(y)$ and for the  $\theta$-dependent anti-neutrino-neutrino scattering (and its conjugate), $g_{\overline{\nu}\nu}(y)$ -- we write:
\begin{equation}
g(y) = \frac{\Gamma_{\nu\nu}}{\Gamma_{\nu}} \left(g_{\nu\nu}(y) + g_{\nu\nu}(1-y)\right) + \frac{\Gamma_{\overline{\nu}\nu}}{\Gamma_{\nu}} \left(g_{\overline{\nu}\nu}(y)+ g_{\overline{\nu}\nu}(1-y)\right)
\end{equation}
where the ratios of scattering rates of $\nu\nu$ and ${\overline{\nu}}\nu$ are  $0.6$ and $0.4$, respectively. We now can rewrite equation \eqref{source2}:
\begin{equation}
S_{\text{ter}}(t,E) =  \int_E^{\infty} g\left(\frac{E}{E'}\right) \frac{\Gamma_{\nu\nu}(t,E') \Phi(t,E')}{E'} dE'
\end{equation}
We can now set up the Boltzman equation which describes the thermal evolution of the differential neutrino flux:
\begin{equation} \label{eqn:Boltzman}
\frac{\partial \Phi}{\partial t} = -2H \Phi +H E \frac{\partial \Phi}{\partial E} + S_{\text{dec}} + S_{\text{ter}} - \Gamma_{\text{tot}} \Phi
\end{equation} 
This partial differential equation can be solved numerically to obtain the present-day differential flux.
\begin{figure}[h]
\begin{center}
\includegraphics[width=\textwidth]{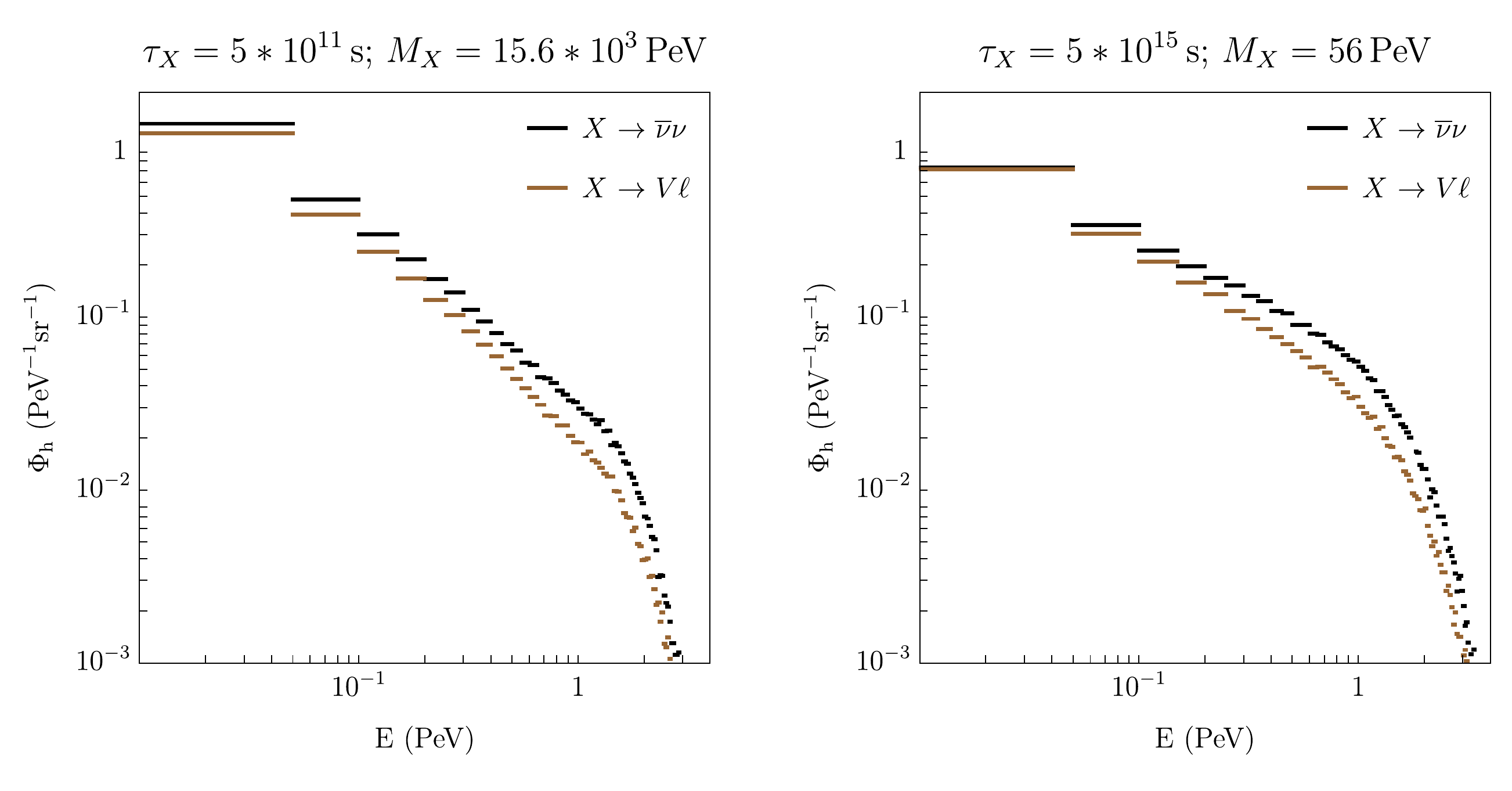}
\caption{$\Phi_{\text{h}}(t_0,E)$ for a heavy decaying relic $X$ for two different lifetimes and two different decay models. The mass of $X$ is set to $M_X = 2.4 \left(1 + z_{\tau_X}\right)$ PeV, the best fit mass of the neutrino spectrum measured at IceCube in the energy range $0.25-4\:{\rm PeV}$ \cite{Kopper:2017zzm}. $z_{\tau_X}$ is the redshift $z$ at the decay lifetime $\tau_X$.}
\label{fig:hist} 
\end{center}
\end{figure}
In our analysis, we implement a propagation code to track the cosmological evolution of individual neutrinos, which is equivalent to solving equation \eqref{eqn:Boltzman} in small time steps. 

For a given lifetime $\tau_X$ we generate events over the appropriate distributions of redshifts, a decaying exponential. The energy spectrum of the injected neutrino is determined by the EW shower. Based on the decay redshift, $z$, we divide the total traveling time of the neutrino into intervals such that the average number of scatterings within the interval is much smaller than one. If a scattering event occurs within a time step, the probability of re-injecting two neutrinos with energy $g(y)$ and $g(1-y)$ is weighted by $\frac{\Gamma_{\nu}}{\Gamma_{\text{tot}}}$. If two neutrinos are re-injected, they undergo the same treatment as the primary injection, starting at redshift $z'$, where the scattering has occurred. This process iteratively continues until the neutrinos either arrive today or scatter into charged leptons or quarks.

The output of the simulation is a histogram $\Phi_{\text{h}}(t_0,E)$, shown in Figure \ref{fig:hist}, which is related to the differential neutrino flux described in equation \eqref{eqn:Boltzman} by dividing by $X$'s number density: 
\begin{equation}
 \Phi_{\text{h}}(t_0,E) \equiv  \Phi(t_0,E)/n_{X,0}(t_0)
\end{equation}
where $\Phi(t_0,E)$ is the solution to \eqref{eqn:Boltzman} at $t=t_0$ and thus accounts for tertiary neutrinos and EW effects. For short lifetimes, including the tertiary neutrinos significantly enhances the flux of lower energy neutrinos, whereas for long lifetimes, these have negligible impact, since almost no scattering occurs. In the limit of negligible tertiary neutrinos, equation \eqref{eqn:Boltzman} can be solved analytically \cite{Ema:2013nda}. 

In our analysis, we only account for neutrino fluxes emerging from extragalactic relic decays.  Extragalactic decays are the only relevant neutrino source for relics with lifetimes $\tau_X \leq 8*10^{16}$s, while galactic decays become important when considering longer lived relics \cite{Ema:2013nda,Cohen:2016uyg,Denton:2017csz}. We leave a detailed investigation of that region of parameter space to future work.

\subsection{Estimating $X$'s Number Density}
We use  $\Phi_{\text{h}}(t_0,E)$ to estimate the number density $n_{X,0}(t_0)$ needed to roughly produce the excess number of events seen in the high energy bins at IceCube \cite{Kopper:2017zzm}. The number of predicted events in this range at the IceCube detector is obtained by integrating over the differential flux times the effective area $A_{\text{eff}}(E)$, which is provided by the IceCube collaboration \cite{Kopper:2017zzm}, and multiplying by the detection time $T$ ($2078\,$days), and solid angle $4 \pi$, as well as the flux velocity $v=c$ to restore SI units.  
\begin{equation} \label{est3}
 N = \int_{E_{\text{min}}}^{E_{\text{max}}} \Phi (t_0,E) A_{\text{eff}}(E) dE * 4 \pi *v * T
\end{equation}
Based of the total number of events ($N_t = 5$) in the range $0.25-4\:{\rm PeV}$ in \cite{Kopper:2017zzm} we estimate the number density $ n_{X,0}(t_0)$ that is needed to produce the observed number of events:
\begin{equation} \label{est}
n_{X,0}(t_0) = \frac{N_{t}}{\int_{E_{\text{min}}}^{E_{\text{max}}} \Phi_h(t_0,E) * A_{\text{eff}}(E) dE * 4 \pi *v* T}
\end{equation}
 
\section{Constraints}
In the following sections we consider different observables that can be used to constrain heavy decaying relics, and how these constraints affect the relic models best suited to generate the PeV neutrinos observed at IceCube.  The summary of our findings appear in Figure \ref{fig:cons}.  The shortest lived relics, those with $\tau_X\leq10^{12}\,\text{s}$, are most strongly constrained by their impact on the abundance of light elements generated during big bang nucleosynthesis (BBN).  Relics with intermediate lifetimes, $10^{12}\,\text{s}< \tau_X\leq 5*10^{15}\,\text{s}$, are most strongly constrained by their impact on the CMB anisotropy power spectrum. Relics with slightly longer lifetimes, $5*10^{15}\,\text{s}<\tau_X\leq 8*10^{16}\,\text{s}$, are most strongly constrained by the $\gamma$-ray spectrum they generate. These constraints all depend on the amount of energy injected into the thermal bath in the form of  electromagnetically interacting (EM) particles. 
In order to explore constraints on our relic models we define $\Xi$,  the EM energy density produced by relic decays divided by the energy density of cold dark matter $\rho_{\text{CDM}}$:
\begin{equation} \label{est2}
 \Xi \equiv f_{\text{int}} \frac{n_{X,0}*M_X}{\rho_{\text{CDM}}}   
\end{equation} 
Here $f_{\text{int}}$ is the fraction of the relic energy density that becomes EM energy and should in principle be redshift-dependent due to rescattering.  However, for the parameter range we are considering, the dominant source of EM energy is from the decay shower where this fraction is largely $M_X$-independent.  We take $M_X = 2.4\, \left(1 + z_{\tau_X}\right)\,$PeV, which gives the best fit mass for the two particular lifetimes shown in Figure \ref{fig:ice}, where $z_{\tau_X}$ is the redshift $z$ at the decay lifetime $\tau_X$. We use this mass as a benchmark for evaluating the constraints for all lifetimes shown in Figure \ref{fig:cons}.   

Based on the results of the EW shower we estimate a conservative lower bound of $f_{\text{int}} = 0.25$ for both decay models. This estimate assumes that about one third of all hadronic energy is electromagnetically interacting, as well as one third of the energy coming from muon and tau decays. This is the number we use for all constraints below.

\subsection{Light Element Abundances} Helium-3 (He$^3$) and Deuterium (D) are produced during Big Bang Nucleosynthesis (BBN) and their measured abundances are in general agreement with the predictions of BBN (see review in \cite{Tanabashi:2018oca}). Decays of heavy relics can initiate EM cascades that interact with the light elements and alter their abundances.  Injected EM particles with energies above $27$ MeV can participate in all of the photodisintegration processes pertinent to producing excess He$^3$ and D by destroying larger nuclei, primarily Helium-4 (He$^4$), as well as those that break He$^3$ and D down into protons \cite{Kawasaki:1994af,Jedamzik:2006xz,Cyburt:2002uv}. Constraints arise from numerically following the evolution of the abundances of all light elements involved in the creation or destruction of He$^3$ and D, and comparing the end predicted abundances to the measured He$^3$ and D abundances \cite{Kawasaki:1994af,Kawasaki:1994sc,Protheroe:1994dt,Jedamzik:2006xz,Poulin:2015opa}. This process, and the resultant constraints on a decaying particle injecting EM energy into the thermal bath, have already been worked out in detail by \cite{Kawasaki:1994af,Kawasaki:1994sc,Protheroe:1994dt,Jedamzik:2006xz,Poulin:2015opa}. We utilize those constraints on the allowed energy density and lifetime of a heavy decaying particle \cite{Poulin:2015opa,Poulin:2016anj}. 

\subsection{CMB Anisotropies}
 EM energy injection by heavy decaying relics with lifetimes in the range $10^{12}\,\text{s}\lesssim\tau_X\lesssim 5*10^{15}\,\text{s}$ can increase the free electron fraction around recombination, thereby distorting the CMB anisotropy power spectrum.  Detailed  constraints have been worked out in \cite{Poulin:2016anj,Slatyer:2016qyl} and we rely heavily on their results, which utilize Monte Carlo Markov chains to calculate the effect of EM energy injection on the CMB anisotropy power spectrum.  This study \cite{Poulin:2016anj} rules out relics that inject enough EM energy at specific redshifts to produce power spectra inconsistent with current measurements.

The injection of EM energy increases the free electron fraction via ionization and collisional excitation.  For relics with lifetimes of $10^{14}\,\text{s}\,\lesssim\tau_X\lesssim 10^{18}\,\text{s}$, the decays enhance the optical depth of the universe after recombination, leading to an additional suppression of the CMB temperature angular power spectra (TT) and polarization power spectra (EE) at small angular scales \cite{Poulin:2016anj}.  Additionally, the increase in the free electron fraction at times between recombination and reionization increases the probability that photons scatter before reionization. This leads to extra polarization, which creates a bump in the  EE spectrum at smaller angular scales than the usual reionization bump \cite{Poulin:2016anj}.  

Relics with  lifetimes of $\sim 10^{13}\,\text{s}$ are the most strongly constrained by the CMB anisotropy.  The EM particles released by relics with lifetimes $\lesssim 10^{13}\,\text{s}$ delay recombination. This widens the last scattering surface, damping the temperature power spectra at small angular scales.  Like the longer lived particles mentioned above, particles with lifetimes $\tau_X \lesssim 10^{13}\,\text{s}$  also generate a bump in the EE spectra to smaller angular scales than the usual reionization bump, though the effect is weaker than that generated by longer lived particles \cite{Poulin:2016anj}.  At times much earlier than $10^{13}\,\text{s}$ , the universe is fully ionized, and injection of electromagnetic particles, which increase the ionization fraction, have little impact. As a result, distortions to the CMB anisotropy spectum are exponentially suppressed for relics with $\tau_X$ much less than $10^{13}\,\text{s}$. At lifetimes $\sim 10^{12}\,\text{s}$, only a fraction of the relics decay late enough to alter the CMB and the constraints from CMB anisotropies become weaker than those that arising from BBN.  

The analysis done by \cite{Poulin:2016anj} only considers the effects of particles with kinetic energies in the range [$10\,$keV, $1\,$TeV], well below the energies of EM particles relevant to our models.  We argue that the bounds also apply to injected EM particles with  $E \geq$ $1$ PeV because, around recombination, EM particles at these energies scatter off the CMB quickly enough to redistribute their energy to many particles with energies below $1\,$ TeV, well within one Hubble time -- energetic photons scatter off CMB photons via pair production extremely efficiently at these energies and redshifts. 
Electrons and positrons scatter off of the CMB through inverse Compton scattering, which while less efficient than pair production at these energies, is still much faster than the Hubble expansion rate for electrons of all energies considered in this paper, as can be verified.


Different injection energies in the range between [$10\,$keV, $1\,$TeV] have different efficiency factors determined by their interactions with the thermal bath, which sets the width of the constraints in \cite{Poulin:2016anj}. To know exactly where within this band our injection energies lie one would have to do a dedicated study. Here, we conservatively apply the least stringent bounds, which correspond to the lowest efficiency of dumping the electromagnetic energy into the thermal bath, noting that a dedicated study for our particular injection energies may improve these bounds by up to a factor of five.   
\begin{figure}[h]
\begin{center}
\includegraphics[width=0.7\textwidth]{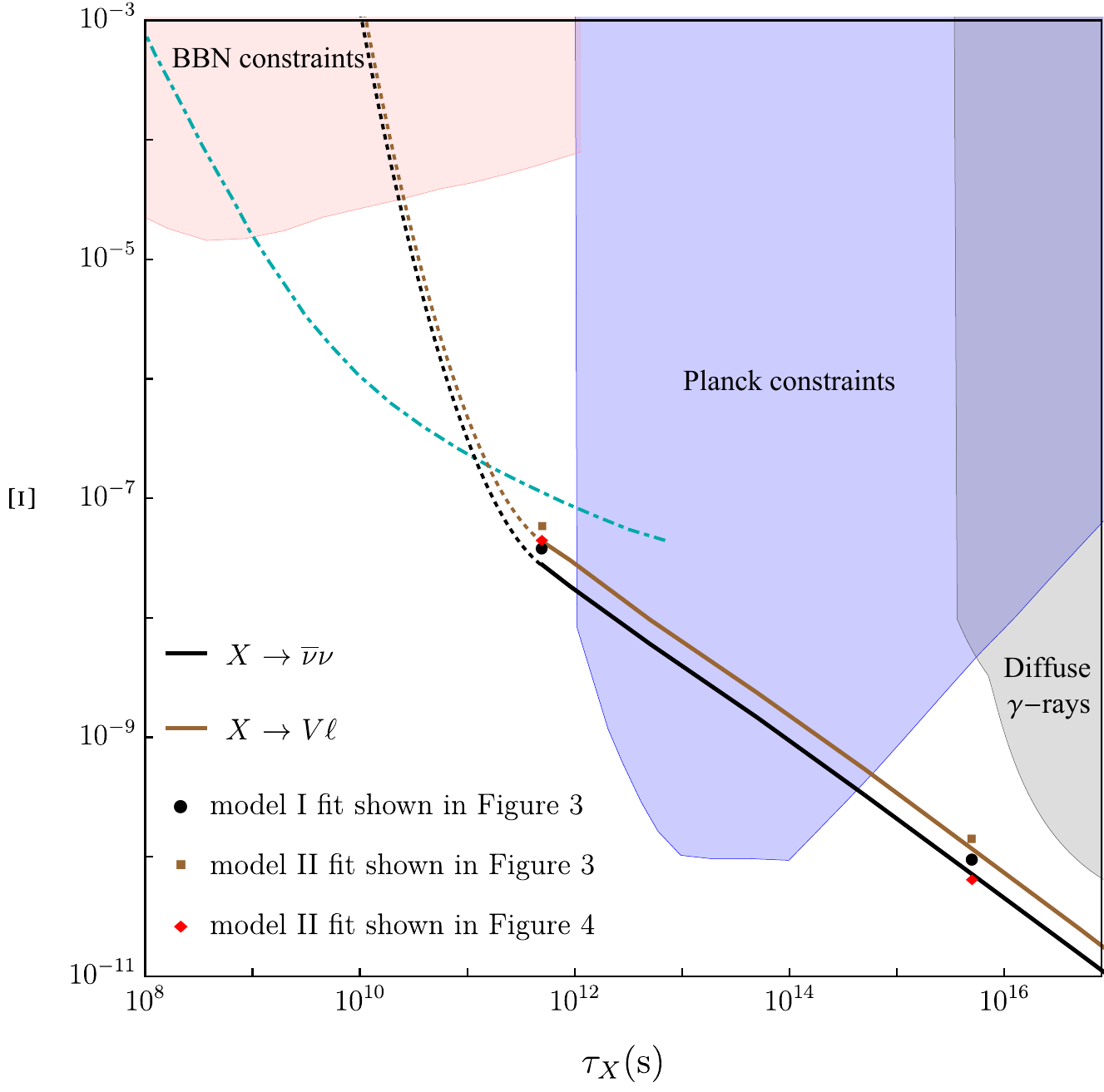}
\caption{Constraints on a wide array of different lifetimes for a heavy decaying relic $X$, releasing EM energy into the thermal bath. All constraints are at $95\%$ confidence level. The light red shaded area is excluded by measurements of light element abundances and their agreement with BBN predictions. The blue shaded area is excluded by bounds from CMB anisotropies. The gray shaded region is excluded by diffuse $\gamma$-ray observations. All of these constraints are for injections of EM energy above some threshold value unique to the constraint and described in their respective sections of this paper.  The cyan line is the forecast from the proposed PIXIE experiment, which could place more stringent bounds from $y$-distortion \cite{Kogut:2011xw}. The black and brown lines indicate the abundance necessary to produce the excess IceCube neutrinos for models I and II, based off equation \eqref{est} and \eqref{est2}, assuming $M_X = 2.4\, \left(1 + z_{\tau_X}\right)\,$PeV. The black and brown (red) markers indicate the data points corresponding to the IceCube spectrum shown in Figure \ref{fig:ice} (Figure \ref{fig:muon}). The dotted lines indicate $M_X = 2.4\, \left(1 + z_{\tau_X}\right)\,$PeV transitioning to an approximation rather than a best fit, as rescattering effects can change the electromagnetic fraction by ${\cal O}(1)$.}
\label{fig:cons} 
\end{center}
\end{figure}

\subsection{$\gamma$-Ray Constraints} 
When the heavy relic decays, the EW shower and decays of the showering products produce energetic photons, electrons, and positrons. These are reprocessed, producing a lower energy $\gamma$-ray distribution, primarily by inverse Compton scattering and pair production \cite{Zdziarski:1989}. The $\gamma$-rays in this reprocessed spectrum lie in the energy range visible to the Fermi telescope, between $0.1$ GeV and $820$ GeV \cite{Ackermann:2014usa}. We derive constraints by requiring that the reprocessed spectra of  heavy relic decays produce a $\gamma$-ray flux that is, in any bin, no more than $2\sigma$ above the flux presented in the Fermi Pass 7 Isotropic Extragalactic Gamma Ray (IGRB) spectrum \cite{Ackermann:2014usa}.

In order to derive the reprocessed $\gamma$-ray spectrum resulting from a heavy relic decay, we follow \cite{Zdziarski:1989,Kribs:1996ac}. Processes by which $\gamma$-rays can lose energy include photoioization, Compton scattering, photon matter pair production, and scattering off of the CMB. In this analysis, we approximate the $\gamma$-ray spectra as if EM particles are only reprocessed by the dominant scattering mechanisms for a particular redshift and energy. We also assume that a photon does not scatter if it has an  optical depth $d_{\tau}<1$.  In this context, the optical depth can roughly be thought of the average number of times a photon scatters as it travels toward the Earth.

For redshifts $0<z\leq 700$, EM particles are reprocessed by initiating cascades with CMB photons through pair production and photon-photon scattering \cite{Zdziarski:1989}.  Pair production is generally more efficient at reprocessing EM particles, except in a small range of energies for $300\leq z\leq 700$, in which photon-photon scattering is more efficient. Photon-photon scattering has a negligible effect on the constraints of relics with $\tau_X\geq 5*10^{15}\,\text{s}$, so we only consider the effect of pair production cascades in this analysis. In pair production cascades, photons pair produce electrons and positrons with CMB photons. The resulting electrons and positrons then upscatter CMB photons by inverse Compton scattering. These two processes continue until the COM energy falls below the pair production threshold. EM particles with energies above the threshold \cite{Zdziarski:1989, Kribs:1996ac}: 
\begin{equation}
 E_{\text{th}}(z) = \frac{m_e^2}{30 \text{ T}(z)}\approx \frac{36 \text{ TeV}}{1+z}   
\end{equation}
have an optical depth $d_{\tau}>1$. Particles with energies below $E_{\text{th}}$ have optical depths $d_{\tau}<1$, in which case we assume they free-stream toward the earth.  At $z>700$, additional scattering processes become important and all EM particles relevant to this analysis thermalize and do not produce any $\gamma$-rays observable today \cite{Zdziarski:1989}.

\begin{figure}
\begin{center}
\includegraphics[width=\textwidth]{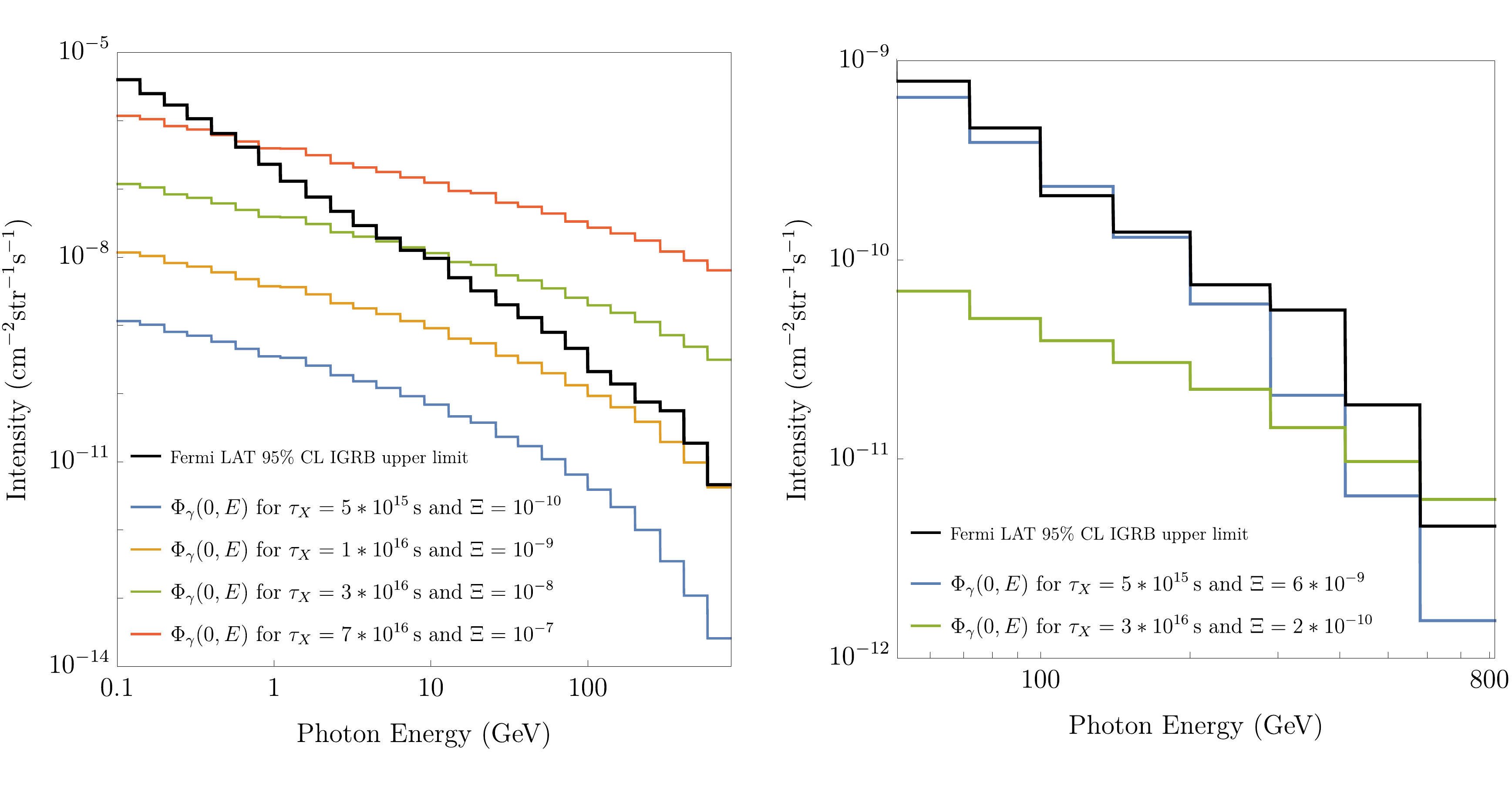}
\caption{Comparison of the Fermi LAT $95\%$ CL IGRB upper limit with the derived diffuse isotropic $\gamma$-ray flux $\Phi_{\gamma}(0,E)$ produced by a decaying heavy relic $X$ for different lifetimes and abundances. The blue line in the left plot corresponds to the black dot in Figure \ref{fig:cons}, which is the abundance necessary to obtain the model I fit shown in Figure \ref{fig:ice}. Since $\Phi_{\gamma}(0,E)$ scales linearly in intensity with $\Xi$ the spectra for the brown dot (model II fit shown in Figure \ref{fig:ice}) and red dot (model II fit shown in Figure \ref{fig:muon}) can be obtained by multiplying the blue line by a factor of $\frac{7}{5}$ and $\frac{2}{3}$, respectively. Shorter lifetimes ($\lessapprox 8*10^{15}\,$s) are most stringently constraint by the energy range between $100$-$140\,$GeV. For longer lifetimes the highest energy range from $580$-$820\,$GeV has the highest constraining power.}
\label{fig:gammarays} 
\end{center}
\end{figure}

Since the particle cascades occur quickly compared to the expansion rate of the universe \cite{Kribs:1996ac}, we define a universal `instantaneously' generated differential $\gamma$-ray cascade spectrum per unit injection energy\footnote{Note that our definition differs from the one given in \cite{Zdziarski:1989}.} $\mathcal{L}(E,z)$, such that $N_{\gamma} =E_{\text{inj}}\int\mathcal{L}(E,z)dE$, where $N_{\gamma}$ is the number of $\gamma$-rays produced when $E_{\text{inj}}$ of energy is injected into the thermal bath at redshift $z$. $\mathcal{L}(E,z)$ is built into the source term in the Boltzmann equation describing the evolution of the differential $\gamma$-ray flux:
\begin{equation} \label{bolz2}
    \frac{\partial\Phi_{\gamma}}{\partial t} = -2H\Phi_{\gamma}+HE\frac{\partial\Phi_{\gamma}}{\partial E}+S_{\gamma}
\end{equation}
where $S_{\gamma}$ is the source term and $H$ is the Hubble parameter.  For a heavy relic, whose decays initiate EM cascades, the source term is:
\begin{equation}
    S_{\gamma}(t,E) = \frac{1}{4\pi}\frac{M_X f_{\text{int}}}{\tau_X}n_{X,0}(t_0)e^{-\frac{t}{\tau_X}}\left(1+z(t)\right)^3 \mathcal{L}(E,z(t))
    \label{gamma}
\end{equation}
 Here we use $M_X f_{\text{int}}$ to denote the total EM energy injected per relic decay\footnote{In defining this source term we assume that all EM particles that result from relic decay are energetic enough to initiate a particle cascade. In general, one would need to consider a source term where the fraction of the relic mass energy that becomes EM particles capable of initiating a cascade depends on $z$.}. Solving the above Boltzmann equation \eqref{bolz2}, gives the diffuse $\gamma$-ray flux for any given $z$. 
\begin{equation}
\label{eqn:Gamma}
    \Phi_{\gamma}(z, E) = \frac{M_X n_{X,0}f_{\text{int}}(1+z)^2}{4\pi\tau_X}\int^{700}_{z}\frac{dz' }{H(z')}\mathcal{L}\left(E\frac{1+z'}{1+z},z'\right)e^{-\frac{t(z')}{\tau_X}}
\end{equation}
For observational purposes, we are interested in the flux at $z$ = 0. We compare the derived diffuse $\gamma$-ray flux today, $\Phi_{\gamma}(0,E)$, for different lifetimes and abundances in Figure \ref{fig:gammarays}.
\begin{equation}
\label{eqn:Gamma}
    \Phi_{\gamma}(0, E) = \Xi* \frac{\rho_{CDM}}{4\pi\tau_X}\int^{700}_{0}\frac{dz' }{H(z')}\mathcal{L}\left(E(1+z'),z'\right)e^{-\frac{t(z')}{\tau_X}}
\end{equation}

$\mathcal{L}(E,z)$ depends on the dominant scattering process for a given redshift.  The cascade spectrum for pair production was numerically calculated by \cite{Zdziarski:1988}.  Here, we use an approximate result only taking into account pair production (as the effects where photon-photon scattering is dominant are negligible):
\begin{equation}
    \mathcal{L}(E_{\gamma},z) = 
    \begin{cases}
    0.767 E_{\text{th}}(z)^{-0.5}E_{\gamma}^{-1.5}, & 0\le E_{\gamma} < 0.04E_{\text{th}}(z)\text{ and }z<700\\
    0.292 E_{\text{th}}(z)^{-0.2}E_{\gamma}^{-1.8}, & 0.04E_{\text{th}}(z)\le E_{\gamma}< E_{\text{th}}(z)\text{ and } z<700\\
    0, & E_{\text{th}}(z)\le E_{\gamma}\text{ or }z\geq700
    \end{cases}
\end{equation}
We derive constraints by comparing $\gamma$-ray spectrum that results from relic decay to the Fermi IGRB spectrum \cite{Ackermann:2014usa}, requiring the predicted relic contribution produces less than a $2\sigma$ contribution in any one bin as illustrated in Figure \ref{fig:gammarays}. Our results are shown in Figure \ref{fig:cons}.

\subsection{Other Constraints}

Spectral distortions to the CMB are often used to constrain the release of EM energy in the early universe \cite{Chluba:2014sma,PhysRevLett.70.2661}. These constraints can be derived by requiring that the decaying relic not produce $\mu$- and $y$- distortions larger than the detection limit of COBE-FIRAS \cite{Fixsen:1996nj}.  These are weaker than the constraints that arise from the light element abundances for the same redshifts, and thus not relevant for this analysis. However, as shown in figure \ref{fig:cons}, the proposed Primordial Inflation Explorer (PIXIE) \cite{Kogut:2011xw}, with projected sensitivities to $\mu$- and $y$-distortions $\sim 1000$x better than those of COBE-FIRAS, could  detect $y$-distortions generated by almost all of the heavy relic models considered in the shorter lifetime parameter space window.

Other works consider constraints on BSM physics from the $21$ cm spin temperature signal \cite{Bowman:2018yin,Kovetz:2018zes,Poulin:2016anj}.  A heavy decaying relic would heat the intergalactic medium, resulting in a positive change to the differential brightness temperature.  We do not consider these constraints in detail in this paper because rough estimates in \cite{Poulin:2016anj} indicate that they are not currently powerful enough to be relevant. However, more data and improvements in the uncertainty of the differential brightness temperature measurement could eventually provide stronger constraints \cite{Bowman:2018yin,Barkana:2018qrx}.

\section{Comparison to IceCube Data}
Figure \ref{fig:cons} shows that there are two windows in which a heavy decaying relic could be the source of the PeV neutrinos observed at IceCube, one with longer lifetimes from $5*10^{14}\,$s to $8*10^{16}\,$s, and one with shorter lifetimes between $7*10^{10}\,$s and $10^{12}\,$s . Here, we show the full neutrino spectrum predicted by the decay of a heavy relic, including neutrinos that result from EW-showering and re-scattering off of the relic neutrino background, for two sample lifetimes within these two allowed ranges. We compare these spectra to six years of IceCube data and we consider data from two different datasets. The first dataset (DS1) includes neutrinos of all flavors that deposited their energy within the detector \cite{Kopper:2017zzm}. The second dataset (DS2) considers six years of IceCube data on upward going muon neutrinos, where the interaction vertex was also allowed to be outside of the detector,  significantly enhancing the effective area \cite{Aartsen:2016xlq}. Both datasets are complementary, and predict roughly the same neutrino fluxes for energies above $3*10^5\,$GeV \cite{Aartsen:2016xlq}. The main focus on our analysis has been on DS1. We still include DS2 in our analysis because it contains the highest energy neutrino event measured to date. The event, which deposited $4.5\,$PeV in the detector, has a $88\%$ probability of being caused by a muon-neutrino, in which case IceCube predicts a reconstructed energy of $7.5\,$ PeV \cite{Aartsen:2016xlq}.  All other anomalous high-energy neutrino events in both data sets have energies below $2.5\,$PeV. We show the best fit for two allowed sample lifetimes for DS1 \cite{Kopper:2017zzm}. We also show the best fit to all of the data by  combining both datasets. We want to stress that the IceCube collaboration has not published a combined measurement, and thus our second set of fits should be taken as purely illustrative.  
\begin{figure}[h]
\begin{center}
\includegraphics[width=\textwidth]{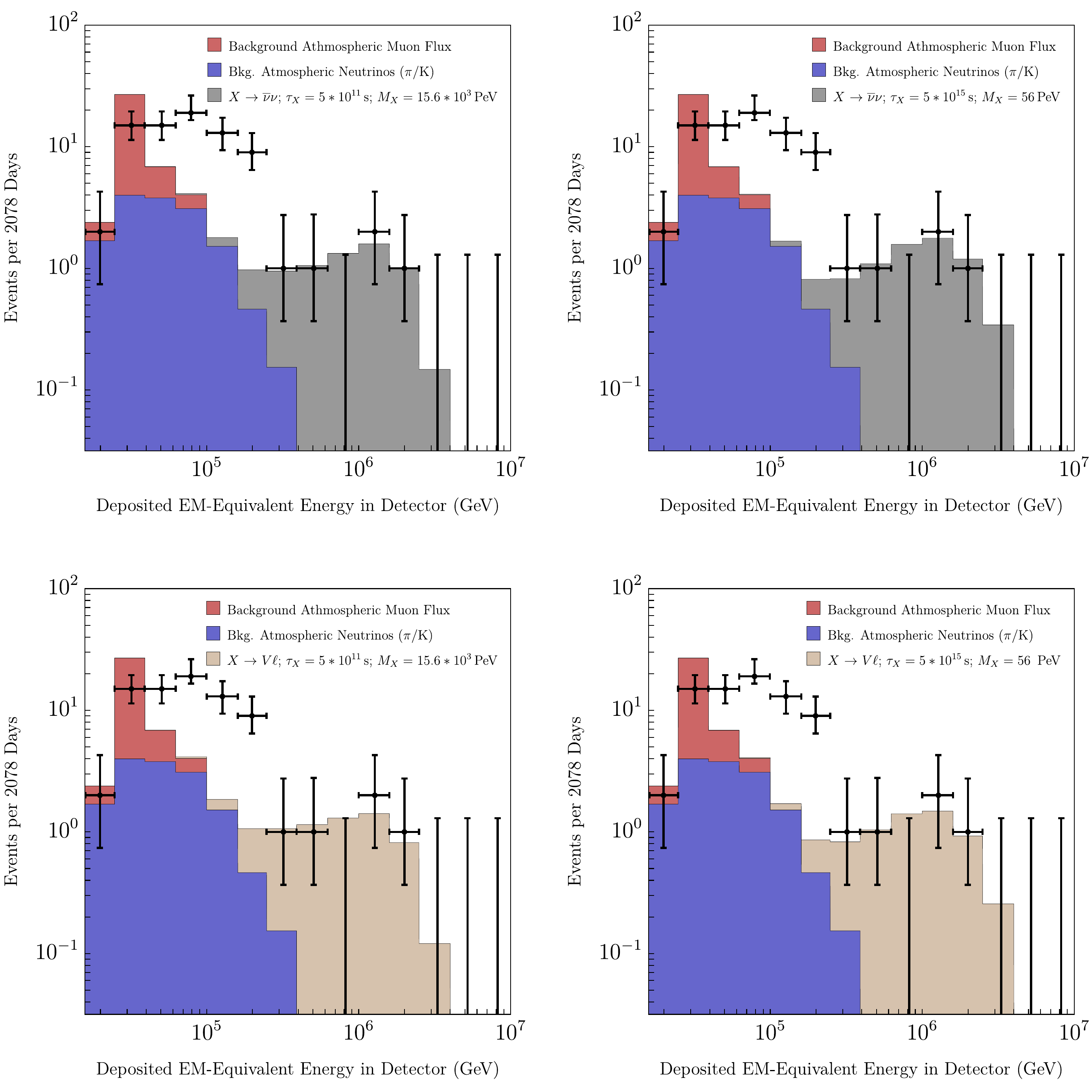}
\caption{ Neutrino spectrum forecast for decay model I and II for two different allowed lifetimes. The displayed spectrum shows the best mass fit for the $0.250-4\,$PeV neutrinos to DS1 for a short sample lifetime $\tau_{X_s} = 5 * 10^{11}\,$s on the left, and a long sample lifetime $\tau_{X_l} = 5*10^{15}\,$s on the right.}
\label{fig:ice} 
\end{center}
\end{figure}
For the following comparison, we choose to fit only to the highest energy events even though there also exists an excess in the lower energy range. We make that choice because a decaying relic cannot comfortably explain both of these excesses at the same time. Other works have considered astrophysical explanations, such as pulsar wind nebulae, Fermi bubbles, and unidentified galactic TeV sources, for this lower energy excess \cite{Fox:2013oza, Lunardini:2013gva, Padovani:2014bha}. One should note that systematic uncertainties and atmospheric backgrounds are much higher in the lower energy range than the higher energy range.  Additionally, DS1 and DS2 are in tension for bins below $ 3*10^5\,$GeV \cite{Aartsen:2016xlq}. The excess of events in the lower energy range is larger in DS1 than in DS2. A better understanding of the tension between the two datasets in this range may be able to give additional insight into the source of the lower energy excess.

\subsection{Dataset 1}
Here, we compare our forecast to DS1 \cite{Kopper:2017zzm}. Figure \ref{fig:ice} shows the neutrino spectrum forecast with the best fit mass to DS1 for two different allowed lifetimes $\tau_{X_s} = 5 *10^{11}\,$s and  $\tau_{X_l} = 5 *10^{15}\,$s. We choose the mass such that the chi-squared is minimized within the range $2.5 *10^5\,$GeV - $4*10^6\,$GeV \cite{Kopper:2017zzm}.  
We can see that for both allowed lifetimes, the spectrum resulting from a heavy decaying relic can reproduce the four highest energy non-zero bins reasonably well. Qualitatively, the spectra do not differ much between the different lifetimes. The shorter lifetime $\tau_{X_s}$ predicts slightly more events between $2.5 *10^5$ GeV - $4*10^5$ GeV, which is an indicator of tertiary neutrinos contributing to the lower tail of the spectrum. While overall there is some contribution to the lower energy bins between $6*10^4\,$GeV - $2.5*10^5\,$GeV, which relieves some of the tension between the expected background and the measurement, it is still an order of magnitude too small to be in agreement with the data.  This suggests that different sources or systematic backgrounds would be needed to explain the excess seen between $5 *10^4$ GeV - $2.5 *10^5$ GeV.

\begin{figure}
\begin{center}
\includegraphics[width=\textwidth]{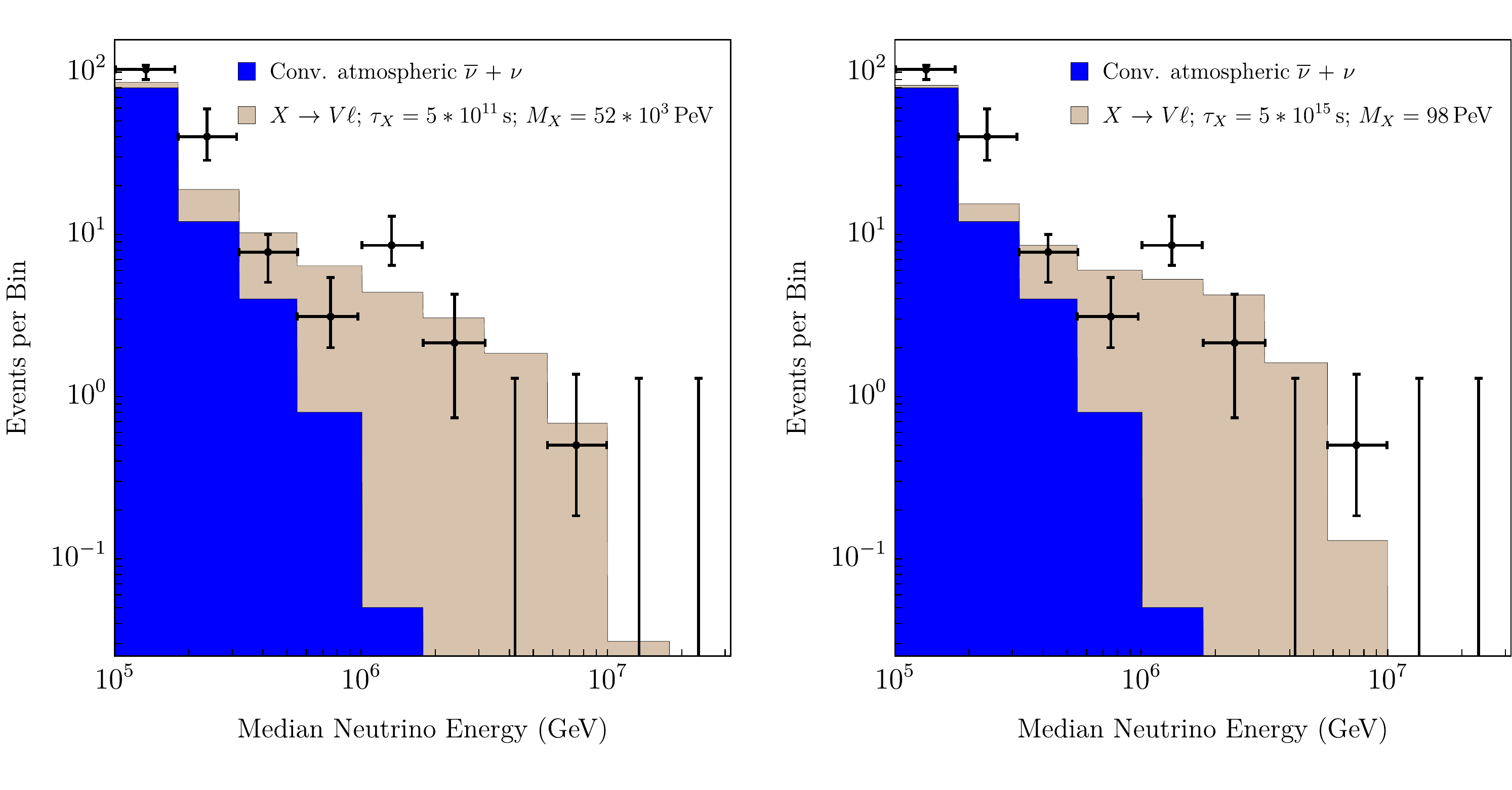}
\caption{ Neutrino spectrum forecast from decay model II ($V \rightarrow V\ell$) for two different allowed lifetimes. The displayed spectrum shows the best mass fit in the range $0.3\,$PeV - $10\,$PeV to the combined dataset for a short sample lifetime $\tau_{X_s} = 5 * 10^{11}\,$s on the left, and a long sample lifetime $\tau_{X_l} = 5*10^{15}\,$s on the right. }
\label{fig:muon} 
\end{center}
\end{figure}

\subsection{Combined Datasets 1 and 2}
 

To combine both datasets, we rearrange equation \eqref{est3} to find the average flux per bin $\Phi_a$ as predicted by the number of events per bin, $N_b$, in DS1:
\begin{equation}
\Phi_a = \frac{N_b}{\int_{E_{\text{min}}}^{E_{\text{max}}} dE \, A_{eff}(E)* 4 \pi *v *T } 
\end{equation}
$E_{\text{min}}$ and $E_{\text{min}}$ correspond to the lower and upper limit in each bin in DS1. We consider all bins between $2.5*10^5\,$GeV - $10^7\,$GeV.
We then calculate how many events per bin, $N_p$, the average flux $\Phi_a$ predicts in DS2:
\begin{equation}
N_p = \int_{E_{\text{min}}'}^{E_{\text{max}}'} dE \, \Phi_a A_{eff}'(E)* 2 \pi *v *T* \eta_f
\end{equation} 
$E_{\text{min}}'$ and $E_{\text{max}}'$ correspond to the lower and upper limit in each bin in DS2. $A_{eff}'(E)$ is the effective detection area for DS2 provided in \cite{Aartsen:2016xlq}. $\eta_f = \frac{1}{3}$ is the flavor efficiency factor, accounting for DS2 only being sensitive to muon-neutrinos.

Combining both datasets shifts the best mass fit from ${M_X} = 2.4 \left(1 + z_{\tau_X} \right)$ PeV to $M_X = 8.\, \left(1 + z_{\tau_{X_s}}\right)\,$PeV and ${M_X} = 4.4\, \left(1 + z_{\tau_{X_l}}\right)\,$PeV for the short ($\tau_{X_s} = 5 *10^{11}\,$s) and long ($\tau_{X_l} = 5 *10^{15}\,$s) lifetimes, respectively. Non-surprisingly, including the higher energy event shifts the mass fits towards higher masses. While for the longer lifetime the spectrum shape still shows the remains of a peak centered around $E_{\text{max}}$, the spectrum for the shorter lifetime does not show this feature. This is due to the spectrum being dominated by tertiary neutrinos, which leads to a power-law shape with a hard cut-off. This effect is more pronounced in the combined dataset fit for $\tau_{X_s}$, because higher $M_X$ enhances the scattering rate off of relic neutrinos. 

\section{Conclusion}
We utilize EW corrections to constrain heavy decaying relic abundances, using measurements impacted by EM energy injection, such as light element abundances during BBN, CMB anisotropies, and diffuse $\gamma$-ray spectra. Beyond the scope of our application, in the future EW corrections may be a useful tool to better constrain BSM physics beyond collider reach, using cosmological and astrophysical data.

We derive a precise forecast of neutrino spectra produced by direct decays from heavy relic particles with lifetimes smaller than $ \tau_X < 8*10^{16}\,$s. Due to our analysis including EW showers and tertiary neutrinos, our forecast accurately captures the shape of the possible spectra, and thus can be used as a powerful discriminant against other astrophysical explanations. This will prove useful as IceCube collects more data. IceCube has plans for a large expansion of its detecting abilities referred to as IceCube Gen 2 \cite{Aartsen:2014njl}. These include plans to increase IceCube's detection area by a factor of $10$, which is expected to improve IceCube's detection sensitivity for neutrinos with energies in the range $10\,$TeV -$1\,$EeV by a factor of 10 \cite{Aartsen:2014njl}. 

 Further, we can expect future experiments to shed more insight into the decaying relics proposed here. PIXIE should be able to detect $y$-distortions from relics with $\tau_X\lesssim5*10^{11}\,\text{s}$. 

Any isotropic, long lived decaying relic heavy enough to generate $1\,$PeV neutrinos will also produce a unique $\gamma$-ray spectrum.
  While Figure \ref{fig:cons} reveals that none of the decaying relics considered in this paper are ruled out by their $\gamma$-ray spectra, this may change as Fermi's detection resolution improves, and as more sensitive $\gamma$-ray telescopes come online.  The latest Fermi data analysis, Pass 8, is far more sensitive to point sources than Pass 7, and additional analysis seem to indicate that much of the IGRB flux derived from Pass 7 may actually be unresolved point sources \cite{DiMauro:2016cbj}. Considering the analysis done by \cite{DiMauro:2016cbj}, we conservatively estimate that at least half of the IGRB flux measured in Pass 7 is actually unresolved point sources.  This would tighten the $\gamma$-ray constraints by at least a factor of 2. However, \cite{DiMauro:2016cbj} contends that the entire IGRB measured in Pass 7 could in principle be explained by unresolved point sources, suggesting that the $\gamma$-ray constraints could become significantly tighter, depending on what fraction of the IGRB is eventually found to be unresolved point sources. These constraints will also improve as $\gamma$-ray telescopes with better point source resolution, such as the High Energy Cosmic Radiation Detection facility (HERD) and the Chernekov Telescope Array (CTA), come online in 2020 with expected 10x more sensitivity than current $\gamma$-ray detectors \cite{Knodlseder:2016pey}.

If most point-source contributions to the IGRB are identified, what remains might be a truly isotropic spectrum from a model such as those described here.  Thus, more IceCube data, paired with improved $\gamma$-ray detection sensitivity, may provide a smoking gun for confirming a heavy decaying relic as source of the IceCube neutrinos and thus physics beyond the standard model.  
 \\    
\appendix
\section{Electroweak Showering} \label{EW shower}
\subsection{Electroweak Splitting Functions}
At energies much larger than the electroweak scale, electroweak radiative corrections have a large impact on decay and scattering processes. This has been explored in the literature with regards to a 100 TeV particle collider \cite{Chen:2016wkt}, and indirect dark matter detection spectra \cite{Ciafaloni:2010ti}. These radiative corrections can be approximated by factorizing the differential cross section (or decay width) into the original $2 \to 2$ ($1 \to 2$) process, times the differential probability that one of the final states will emit an additional gauge boson (or split into two different particles altogether).

 At very high energies there will be more splittings, which requires a summation for a full treatment. However, the majority of the higher order splittings are soft, which means they only carry a small fraction of the total energy. To compare our prediction to the spectrum at IceCube we are only interested in neutrinos within two orders of magnitude of the highest energy neutrinos. This allows us to use a cutoff above which the splitting probability does not exceed $1$. In our EW shower we only consider 'hard' first order splittings, in which the gauge boson carries more than $10^{-2}$ of the maximum possible energy. This treatment captures how the resulting spectrum today is affected by the additional particles produced by a decay.  However, at these high COM energies, many soft $W$'s can populate the final state, which can turn a charged particle into a neutral one and vice versa. Therefore at high energies we keep track of all leptons and scalars (for the Higgs and the longitudinal components of the gauge bosons), and do an isospin average in the end. 
 
We use the following splitting functions in the implementation of the EW shower. Here we follow the notation in \cite{Ciafaloni:2010ti}. $D_{A \to B} (x)$ gives the differential probability that a particle A turns into another particle B, with fraction $x$ of the initial energy. Equations \eqref{startS}-\eqref{endS} show the splitting functions for scalars such as the Higgs $h$, and the longitudinal components of the gauge bosons, $W_L$ and $Z_L$. Equations \eqref{startL}-\eqref{endL} show the splitting functions for fermions. All couplings are renormalized. $L(x)$ and $l$ below are the universal kinematical functions \cite{Ciafaloni:2010ti}:
\begin{equation}
L(x) = \ln \frac{s x^2}{4 M_V^2} + 2 \ln \left(1+\sqrt{1-\frac{4 m_V^2}{s x^2}} \right)
\end{equation}
\begin{equation}
l = \ln \frac{s }{ M_V^2} 
\end{equation}

\subsection*{Splitting Functions for $h/Z_L$}
In the following equations $h$ may be replaced with $Z_L$. In the high-energy limit $h$ and $Z_L$ are not distinguishable, which is why they have the same splitting functions. 
\begin{equation} \label{startS}
D_{h \to W_T} (x) = \frac{\alpha_2}{2 \pi} \frac{1-x}{x} L(x)
\end{equation}
\begin{equation}
D_{h \to Z_T} (x) = \frac{\alpha_2 c_w^2}{ \pi} \frac{1-x}{x} L(x)
\end{equation} 
\begin{equation} \label{topsplit}
D_{h \to t} (x) = \frac{3\alpha_t}{2 \pi} l
\end{equation}
Notice that the initial particle spin stays the same when emitting a gauge boson. Here for example we start out with a Higgs $H$, which can emit a $W_T$, which turns the Higgs into a $W_L$, or it can emit a $Z_L$, in which case it remains a Higgs. In either case the mother-particle, which carries the majority of the energy after the splitting, remains a scalar. The Higgs can also split into two top quarks, in which case neither of them have a higher probability of carrying the majority of the energy. This can be seen by \ref{topsplit} being independent of $x$. (Splittings into other quarks and leptons are negligible because their yukawa couplings are small.)
\subsection*{Splitting functions for $W_L$}
\begin{equation}
D_{W_L \to W_T} (x) = \frac{\alpha_2}{2 \pi} \frac{1-x}{x} L(x)
\end{equation}
\begin{equation}
D_{W_L \to Z_T} (x) = \frac{\alpha_2}{ \pi} \frac{(s_w^2-\frac{1}{2})^2}{c_w^2} \frac{1-x}{x} L(x)
\end{equation} 
\begin{equation}
D_{W_L \to \gamma} (x) = \frac{\alpha_{EM}}{ \pi} \frac{1-x}{x} L(x)
\end{equation}
\begin{equation} \label{endS}
D_{W_L \to t} (x) = \frac{3\alpha_t}{4 \pi} l
\end{equation}
\subsection*{Splitting functions for fermions}
Here are the splitting functions for a charged fermion: 
\begin{equation} \label{startL}
D_{f \to W_T} (x) = \frac{\alpha_2}{2 \pi} \frac{1}{2} \frac{1+(1-x)^2}{x} L(x)
\end{equation}
\begin{equation}
D_{f \to Z_T} (x) = \frac{\alpha_2}{2 \pi} \frac{1}{4 c_w^2} \frac{1+(1-x)^2}{x} L(x)
\end{equation}
\begin{equation}
D_{f \to \gamma} (x) = \frac{\alpha_{EM}}{2 \pi} \frac{1}{4 c_w^2} \frac{1+(1-x)^2}{x} L(x)
\end{equation}
Here are the splitting functions for a neutral fermion: 
\begin{equation}
D_{f \to W_T} (x) = \frac{\alpha_2}{2 \pi} \frac{1}{2} \frac{1+(1-x)^2}{x} L(x)
\end{equation}
\begin{equation} \label{endL}
D_{f \to Z_T} (x) = \frac{\alpha_2}{2 \pi} \frac{(s_w^2-\frac{1}{2})^2}{c_w^2} L(x)
\end{equation}
\subsection{Description of Included Processes}

In model I, each decay produces two leptons. In model II, each decay of a heavy $X$-particle produces one scalar and one lepton. We consider one hard splitting off of both daughter particles. We decay all top quarks, keeping track of all gauge bosons. We combine the energy spectrum of the primary lepton with subsequent decays from any gauge bosons ($V_L$ and $V_T$) to secondary leptons. We consider only direct leptonic decays, as neutrinos resulting from hadronic decays are much less likely to be energetic enough to be above our set threshold of $x> 0.01$. Included gauge bosons come from the primary scalar, radiation off of either leptons or scalars, and subsequent decays from top quarks to $W$'s.

 The total lepton spectrum $f_{\text{tot}}(x)$ is the combination of the primary and secondary lepton spectrum. $f_{\text{tot}}(x)$ is the probability distribution of producing a lepton with fraction $x$ of $\frac{M_X}{2}$. Since we have to average over charged and neutral leptons due to the possibility of soft $W$-emission, the probability distribution of a neutrino with energy fraction $x$ of $\frac{M_X}{2}$ is given by $\frac{1}{2}f_{\text{tot}}(x)$. The other half of $f_{\text{tot}}(x)$ results in charged leptons: electrons, muons, and taus. While electrons are stable, muons and taus decay further before interacting with the thermal bath. 
 
\begin{figure}
\begin{center}
\includegraphics[width=\textwidth]{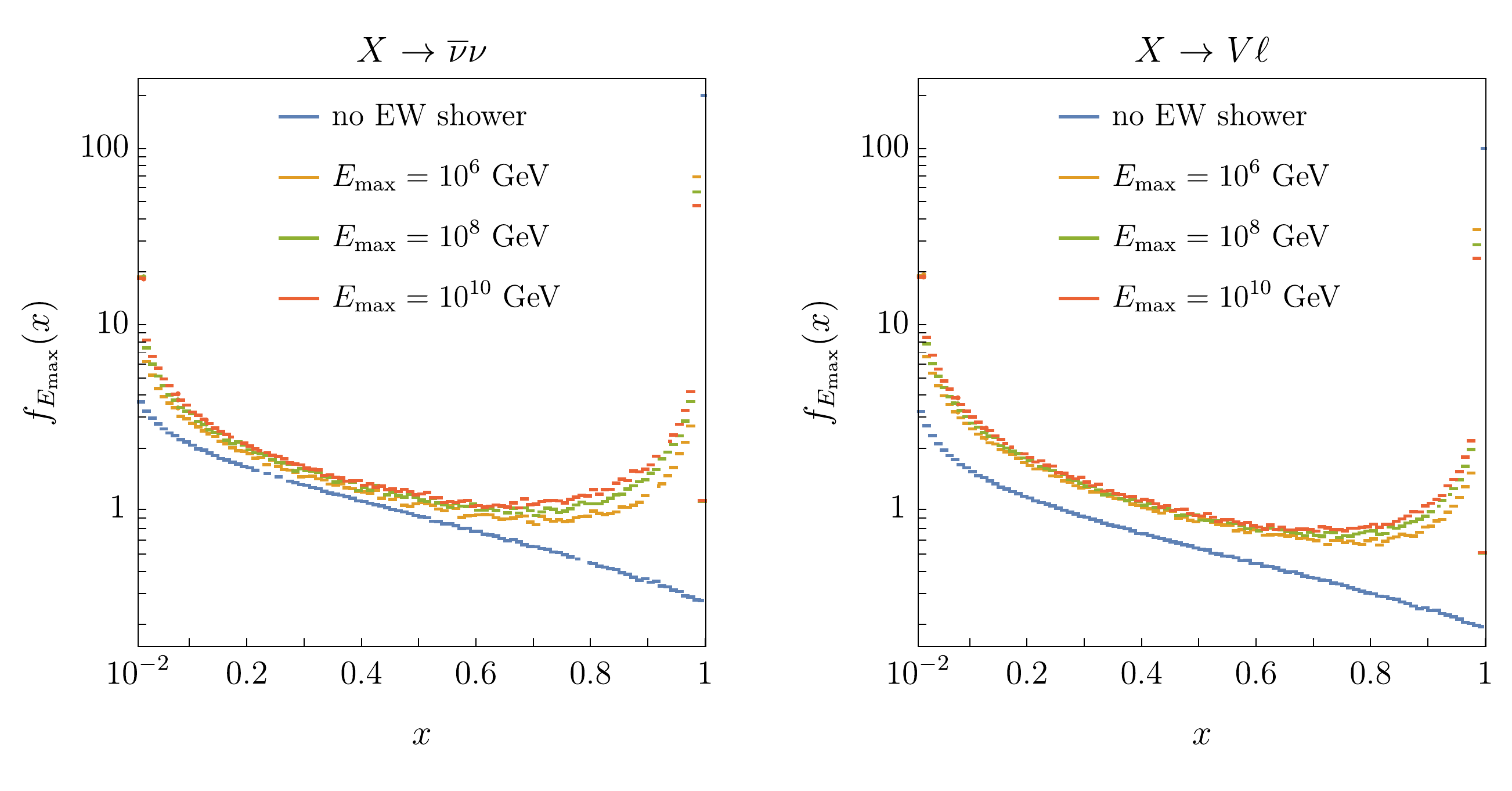}
\caption{ The final neutrino spectrum of decay model I and II considering EW showers at different energies. For comparison the spectrum without including EW showering is shown as well. The final spectrum includes decays to neutrinos from any gauge bosons, taus and muons produced in the EW shower or in the primary decay. We can see that for higher COM energies the peak decreases, which demonstrates how more energy is distributed to EW radiation.       }
\label{fig:EWshower} 
\end{center}
\end{figure}

 Neutrinos from primary muon and tau decays will also contribute to the measured spectrum today. We assume an isotropic three-body decay, and decay all muons into three particles, two of which contribute to the neutrino spectrum. The tau-decays are more subtle as there is a greater variety of possible final states. We treat the leptonic tau decays in the same manner as the muon decays. We also include other tau-decays with up to three particles in the final state and add the resulting neutrinos to the spectrum, without further decaying any resulting mesons. The final neutrino spectrum, which is shown in Figure \ref{fig:EWshower}, is denoted by $f_{E_{\text{max}}}(x)$. 
 
  Neglecting hadronic decays may slightly underestimate the low energy tail of our distribution. In the future, it may be worth integrating an EW-shower formalism with a hadronic shower. For our purposes, the accuracy of the high energy tail of the neutrino distribution is most important, to which hadronic decays will not significantly contribute. 
 
\acknowledgments

We acknowledge the support of NSF grant No. PHY-1818899.  We thank Surjeet Rajendran for collaboration in early stages of the project.  We thank Vivian Poulin and Brock Tweedie for extensive discussions and comments. We also thank Kim Boddy, Junmou Chen, Stephen Mrenna, Ely Kovetz, Kirill Melnikov, Tommi Tenkanen, and J{\"o}ran Stettner for discussions.

\bibliographystyle{JHEP}
\bibliography{ourbib}

\end{document}